\documentclass[aps,prapplied,twocolumn,superscriptaddress,longbibliography]{revtex4-1}

\usepackage{pdfpages}
\usepackage{amsmath, amssymb, braket}
\usepackage{graphicx, hyperref}

\makeatletter
\AtBeginDocument{\let\LS@rot\@undefined}
\makeatother

\begin{document}

\title{In-Plane Ferroelectric Tunnel Junction}

\author{Huitao Shen}
\affiliation{Department of Physics, Massachusetts Institute of Technology, Cambridge, Massachusetts 02139, USA}
\author{Junwei Liu}
\affiliation{Department of Physics, Hong Kong University of Science and Technology, Clear Water Bay, Hong Kong, China}
\author{Kai Chang}
\affiliation{Max-Planck Institute of Microstructure Physics, Weinberg 2, 06120 Halle (Saale), Germany}
\author{Liang Fu}
\affiliation{Department of Physics, Massachusetts Institute of Technology, Cambridge, Massachusetts 02139, USA}

\begin{abstract}
The ferroelectric material is an important platform to realize non-volatile memories. So far, existing ferroelectric memory devices utilize out-of-plane polarization in ferroelectric thin films. In this paper, we propose a new type of random-access memory (RAM) based on ferroelectric thin films with the in-plane polarization called ``in-plane ferroelectric tunnel junction''. Apart from non-volatility, lower power usage and faster writing operation compared with traditional dynamic RAMs, our proposal has the advantage of faster reading operation and non-destructive reading process, thus overcomes the write-after-read problem that widely exists in current ferroelectric RAMs. The recent discovered room-temperature ferroelectric IV-VI semiconductor thin films is a promising material platform to realize our proposal.
\end{abstract}

\maketitle

To meet the daily increasing demands of modern electronic devices, especially of portable devices, memories with low energy consumption and high performance are highly desired. The current commercial dynamic random-access memories (DRAM) are volatile, which consume a large amount of energy to refresh the stored data in order to prevent leakage from the capacitor. To reduce the energy consumption, a non-volatile memory might be the ultimate solution \cite{Scott2007, Kim2011}.

The ferroelectric material has been proposed to be an ideal candidate for non-volatile memories due to its electric switchable bistable ground states since 1952 \cite{Buck1952}, and ferroelectricity based non-volatile memories have been developed rapidly in the past several decades \cite{Scott1400,scott2013ferroelectric}.

Depending on the readout mechanism, ferroelectric non-volatile memories can be roughly classified into two generations. The first generation of ferroelectric RAM (FeRAM) uses polarized charges in the ferroelectric capacitor to represent the data \cite{5940,48273,ishiwara2004ferroelectric}. As a result, discharging the capacitor to measure the polarized charge destroys the stored data, and the capacitor needs to be recharged after the reading operation. Limited by the destructive reading process, the ferroelectric size effects \cite{LI1996341,Lichtensteiger2007} and various practical issues such as fatigue \cite{Tagantsev2001} and imprint \cite{Kang2003}, the market of FeRAM remains relatively small. 

To overcome the destructive readout problem, the second generation of ferroelectric tunnel junction (FTJ) is proposed to probe the ferroelectric polarization using the tunneling electroresistance effect \cite{PhysRevLett.94.246802,PhysRevB.72.125341,Tsymbal181}. The basic structure of the FTJ is a metal-ferroelectric-metal junction, where the tunneling potential barrier is determined by the out-of-plane polarization in the ferroelectric layer. In this way, the FTJ realizes bistable resistance states. The major challenge of realizing FTJ is to fabricate ultrathin ferroelectric films so that the tunneling current surpasses the threshold of peripheral amplifiers. The depolarization field induced by the out-of-plane polarization dramatically suppresses the ferroelectric critical temperature or even destroys the ferroelectricity when the films are too thin \cite{Fong1650,PhysRevLett.96.127601,PhysRevLett.103.177601,PhysRevB.81.064105}. 


In this work, we propose a new type of ferroelectric memory which we call ``in-plane ferroelectric tunnel junction''. Different from FeRAM or FTJ, which employs bistable states of out-of-plane ferroelectric polarization to represent ``ON'' and ``OFF'', our proposal is based on the in-plane polarization of ferroelectric thin films. Due to the insufficient screening in two dimensions (2D), the in-plane polarization could induce strong band bending around the edge or the ferroelectric domain wall. Depending on the polarization direction, the upward/downward band bending could be used to represent the ON/OFF state. By measuring the out-of-plane tunneling current through ferroelectric thin films, the bending direction can be detected and hence the stored information can be read non-destructively. Moreover, our design enjoys great tunability. By choosing proper layer sizes and the insulator layer band gap, the tunneling current and the ON/OFF current ratio can be tuned simultaneously. In principle, all ferroelectric thin films with in-plane polarization can be used as the material platform to realize our proposal \cite{Lee2001, doi:10.1063/1.2136228, Matzen2014, Wang2016,1347-4065-41-8R-5288,Chang2016,doi:10.1002/adma.201804428,Ding2017,Zheng2018,doi:10.1021/acs.nanolett.7b04852,doi:10.1021/acs.nanolett.8b02688}. In particular, the recently discovered room-temperature IV-VI semiconductor thin films with robust in-plane polarization is an ideal candidate \cite{Chang2016,doi:10.1002/adma.201804428}.

The paper is organized as follows. We first introduce in-plane ferroelectric polarizations and the induced robust band bending. Then we demonstrate the device design and explain its reading and writing mechanism in detail. The demonstration is supported by the quantum mechanical tunneling current simulation. Finally, we discuss the advantage of our design over conventional ferroelectricity based memories. 

\section{In-plane polarization} 
Ferroelectricity as a symmetry-breaking state is generally destabilized by the finite-size effect. The out-of-plane ferroelectric polarization is found in perovskite ultrathin films, in which imperfect charge screening, substrate strain and chemical bonding play important roles in stabilizing ferroelectricity \cite{doi:10.1063/1.1662770,Junquera2003,PhysRevB.70.104108,PhysRevB.72.020101,PhysRevLett.102.107601,PhysRevB.90.184107}. As already mentioned in the introduction, the critical temperature of these perovskite ferroelectric materials decreases with the film thickness. On the other hand, in-plane polarization in perovskite thin films \cite{Lee2001, doi:10.1063/1.2136228, Matzen2014, Wang2016} and even liquid crystals \cite{1347-4065-41-8R-5288} are studied. Although it is predicted that in-plane polarization survives in the 2D limit \cite{PhysRevB.81.064105}, currently it is hard to prepare free-standing 2D perovskite ferroelectrics. 

Surprisingly, the recent discovered in-plane polarization in 2D ferroelectrics is enhanced instead of reduced in thin films \cite{Chang2016,Ding2017,Zheng2018,doi:10.1021/acs.nanolett.7b04852,doi:10.1021/acs.nanolett.8b02688}. For example in SnTe, compared with the bulk ferroelectric transition temperature of $98\mathrm{K}$, the one monolayer (ML) thin film has a critical temperature of $270\mathrm{K}$, and thicker films with 3 ML show robust spontaneous polarization even at room temperature \cite{Chang2016,doi:10.1002/adma.201804428}. Moreover, the weak van der Waals interlayer coupling enables more freedom in device design as in principle 2D materials can be stacked freely without the constraint of lattice mismatching.

\begin{figure}[tbp]
\centering
\includegraphics[width=\columnwidth]{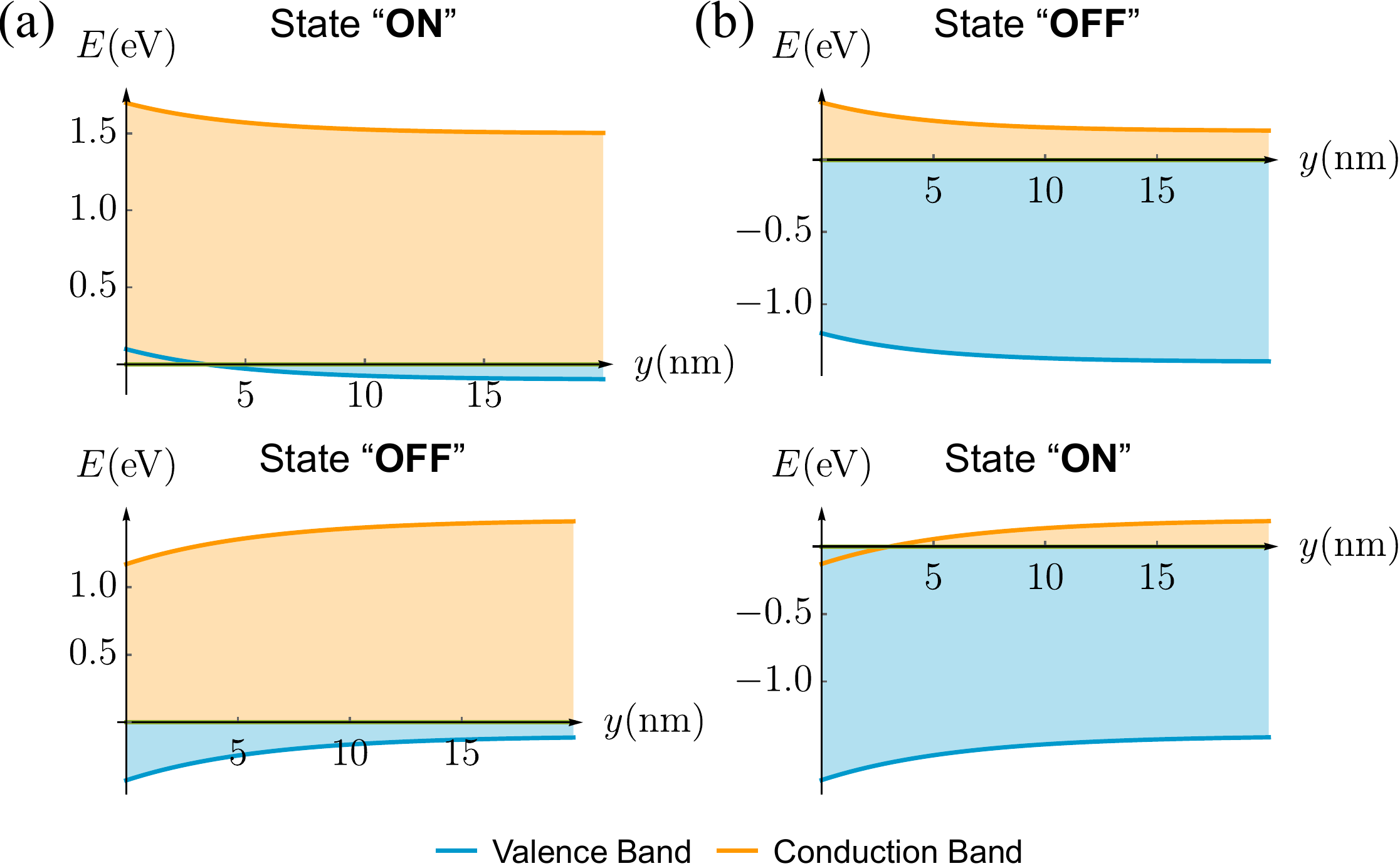}
\caption{Schematic of upward (top) and downward (bottom) band binding near the material edge, for chemical potential near the valence band (a) and conduction band (b). Fermi energy of the electrode is set to be $ E=0 $. 
}
\label{fig:bending}
\end{figure}

An important signature of the in-plane polarization is the band bending near the interface such as the material edge \cite{Chang2016} or the ferroelectric domain wall \cite{Mundy2017}. Taking material edge as an example, without screening, the bound charges induced by the in-plane ferroelectric polarization $ \sigma_b=\mathbf{P}\cdot\mathbf{n} $ are of opposite signs at the two boundaries, where $ \mathbf{P} $ is the polarization vector and $ \mathbf{n} $ is the normal vector of the boundary. The resulting electric field leads to a linear band tilting in 3D and a logarithmic one in 2D, where the energy decreases from the negatively charged boundary to the positively charged boundary. When free charge carriers are present, which could be contributed from the substrate or the defects in the ferroelectric material, the screening effect cancels the boundary charge so that only the band bending near the boundary remains. In the following, we consider ferroelectric materials with large band gap so that the free charge carriers are from the metallic substrate. Due to the insufficient screening in 2D, the band bending can extend quite a region (typically several nanometers) near the interface. For example, the scanning tunneling microscopy measured band bending near the boundary of 1 ML SnTe film can be fitted nicely by an exponential function $ V=\alpha e^{-y/\lambda}+c $ \cite{Chang2016}, which is sketched in Fig.~\ref{fig:bending}. SnTe thin films with odd number of monolayers all share similar band bending profile. We emphasize that this band bending near the interface generally exists for all ferroelectric thin films with in-plane polarization, and does not depend crucially on material details. 

\begin{figure}[tbp]
\centering
\includegraphics[width=.95\columnwidth]{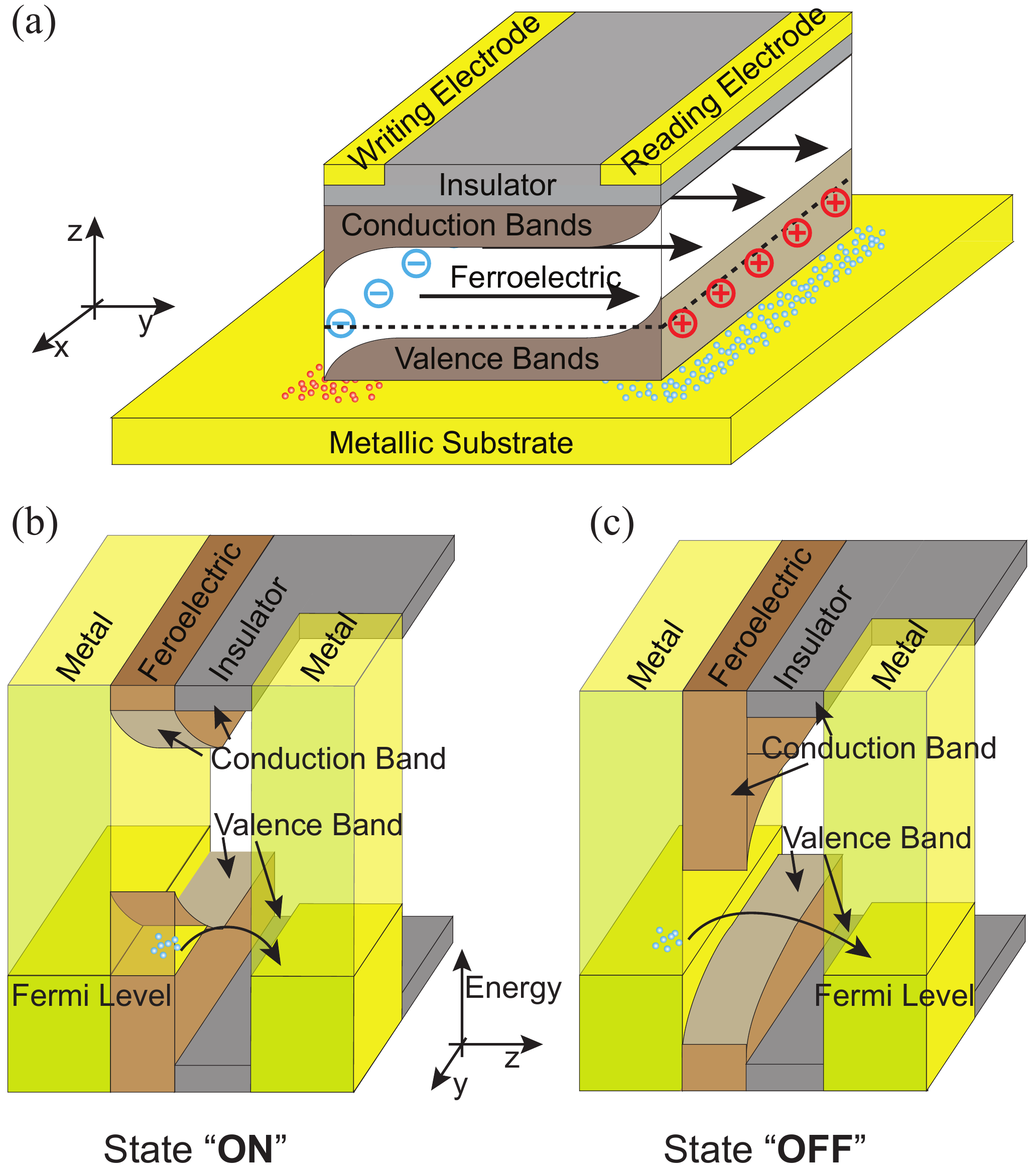}
\caption{Schematic of the device (a) and the band diagram of state ``ON'' (b) and ``OFF'' (c). Here the chemical potential is near the valence band, which corresponds to the left column in Fig.~\ref{fig:bending}. The in-plane polarization induces bound charges and thus robust band bending around the boundary. Depending on the polarization direction, one of the edge could be conducting (b) or insulating (c), which will change the effective potential barrier between the reading electrode and the substrate, therefore realizing bistable states with different tunneling currents. In (a), the large circles with ``$+$/$-$'' represent positive/negative bound charges induced by the ferroelectric polarization. The small red/blue dots represent positive/negative screening charges from the metallic substrate. The dashed line labels the Fermi level. 
}
\label{fig:device}
\end{figure}

\section{Device Design}
The robust band bending induced by the in-plane polarization motivates us to propose a new type of non-volatile memory. The schematic of the device is shown in Fig.~\ref{fig:device}(a). The core of the design is a ferroelectric thin film sandwiched by a metallic substrate and a wide-band-gap insulator. The writing electrode and the reading electrode are deposited at two different edges of the top insulator that is (mostly) parallel to the polarization direction. In the figure, the in-plane polarization is assumed to be along the $ +y $ direction, which will induce opposite net charges at different boundaries. Depending on the polarization direction ($ +y $ or $ -y $), the band bending near the reading electrode could be upward or downward, whose mechanism is already discussed in the previous paragraph. The band diagram near one of the electrode is shown in Fig.~\ref{fig:device}(b) and Fig.~\ref{fig:device}(c), where the chemical potential is set to be near the valence band.

To write the information or manipulate the polarization direction, one can apply a writing voltage $ \pm V_{\rm W} $ on the writing electrode with reference to the reading electrode to generate an in-plane electric field across the ferroelectric thin film, thus forcing the polarization along $ \pm y $ direction regardless of the initial polarization direction. 
The stored information is represented by the $ \pm y $ polarization direction, which can be seen as the ``0/1'' bit. The polarization persists a longer duration even after the write voltage is turned off, and in this way the information storage is non-volatile. The time cost of the writing operation is determined by the polarization switching time, which in turn is determined by the applied writing voltage $ V_{\rm W} $. For example, the coercive field in GeTe thin films is reported to be $ E_c=0.206\mathrm{V/nm} $, which translates to around $ V_c=20.6\mathrm{V} $ for writing and reading electrodes separated by $ 100\mathrm{nm} $ \cite{doi:10.1063/1.4996171}. For $ V_{\rm W}<V_c $, which is usually the case in a realistic device, the switching time is determined by the domain wall dynamics, which is typically several hundreds of picoseconds \cite{doi:10.1063/1.1722712,doi:10.1063/1.1644917,1347-4065-43-5S-2818}.

To read the information or measure the polarization direction, one can apply a reading voltage $ V_{\rm R} $ on the reading electrode with reference to the metallic substrate and measure the tunneling current. The tunneling current depends on the band bending and hence the polarization direction. More specifically, the charge carriers can be holes or electrons if the chemical potential is near the valence or the conduction band respectively. In the case of hole charge carriers, the tunneling current in the ON state $ I_{\rm ON} $, where the band bends upward near the reading electrode, is significantly larger than that in the OFF state $ I_{\rm OFF} $, where the band bends downward. This is illustrated in Fig.~\ref{fig:bending}(a) and Fig.~\ref{fig:bending}(b)(c). In the  case of electron charge carriers, the downward bending represents the ON state and the upward bending represents the OFF state, as is shown in Fig.~\ref{fig:bending}(b). Since the electric field generated by the reading voltage is perpendicular to the polarization, the reading process is non-destructive. 
Note that no capacitor discharge is involved in this process, the time cost of the reading operation is almost only determined by the peripheral current measurement device. 

\section{Tunneling Current}
We now turn to a detailed study of the tunneling electroresistance effect between the metallic substrate and the reading electrode. Without loss of generality, we assume the Fermi level of the metal is close to the valence band of the ferroelectric film. If the band bending is upward and strong enough, the valence band edge would be higher than the Fermi level of the metal, making the ferroelectric thin film conducting. In this case (ON state), the tunneling happens between the ferroelectric thin film and the reading electrode (Fig.~\ref{fig:device}(b)). It is worth noting that the above discussion also works for the scenario of downward band bending if the Fermi level in the metal is close to the conduction band of the ferroelectric film, and a parallel computation is presented in \cite{SM}. On the other hand (OFF state), the downward band bending makes the ferroelectric thin film insulating. The tunneling then happens between the metallic substrate and the reading electrode (Fig.~\ref{fig:device}(c)). In this way, the threshold voltage for the ON state is determined by the band gap and the thickness of the insulator; while the ON/OFF ratio $ I_{\rm ON}/I_{\rm OFF} $ is determined by the band bending and the thickness of the ferroelectric thin film.

To make the above intuitive argument more concrete, we compute the tunneling current in a metal-ferroelectric-insulator-metal junction using two-terminal Landauer's formula 
\begin{equation}
I=\frac{2e}{h}\int_{-\infty}^{\infty}T(E)\left[f_{\rm L}(E)-f_{\rm R}(E)\right]dE=\frac{2e}{h}\int_{0}^{U}T(E)dE, \label{eq:lan}
\end{equation}
where $ T(E) $ is the transmission probability and $ f_{\rm L/R}(E)=\left[e^{-(E-\mu_{\rm L/R})/k_BT}+1\right]^{-1} $ is the Fermi-Dirac distribution function. $ \mu_{\rm L} $ and $ \mu_{\rm R} $ are the chemical potential of the left and the right electrode respectively. In the following, we always set $ \mu_{\rm L}=0 $ as the reference. At zero temperature, $ f_{L/R}(E) $ becomes the step function and Eq.~\eqref{eq:lan} reduces to its final form, where $ U\equiv \mu_{\rm R}-\mu_{\rm L} $ is the voltage bias. 

The geometry of the system is taken to be the same as the device design in Fig.~\ref{fig:device}, where from the $-z$ to $+z$ there are in order: the left metal electrode (substrate), the ferroelectric, the insulator, the right metal electrode (reading electrode). 
The dimension is $ X\times Y \times Z $. Here $ Z=d_{\rm FE}+d_{\rm I} $, which is the thickness of the ferroelectric and the insulator film respectively.

Inside the junction, the electrons and the holes are governed by the Schr\"odinger equation
\begin{equation}
\left[-\frac{\hbar^2\nabla^2}{2m^*}+V(x,y,z)\right]\psi=E\psi. \label{eq:schro}
\end{equation}
The potential barrier of the ferroelectric $ V(x,y,0\leq z<d_{\rm FE}) $ is modeled by the fitted potential of 1 ML SnTe thin film \cite{Chang2016,SnTe}. 
The potential of the insulator $ V(x,y,d_{\rm FE}\leq z<d_{\rm FE}+d_{\rm I}) $ is modeled by a square potential of monolayer h-BN \cite{Kubota932}. All calculation details including parameters can be found in \cite{SM}. 

\begin{figure}[tbp]
\centering
\includegraphics[width=1.0\columnwidth]{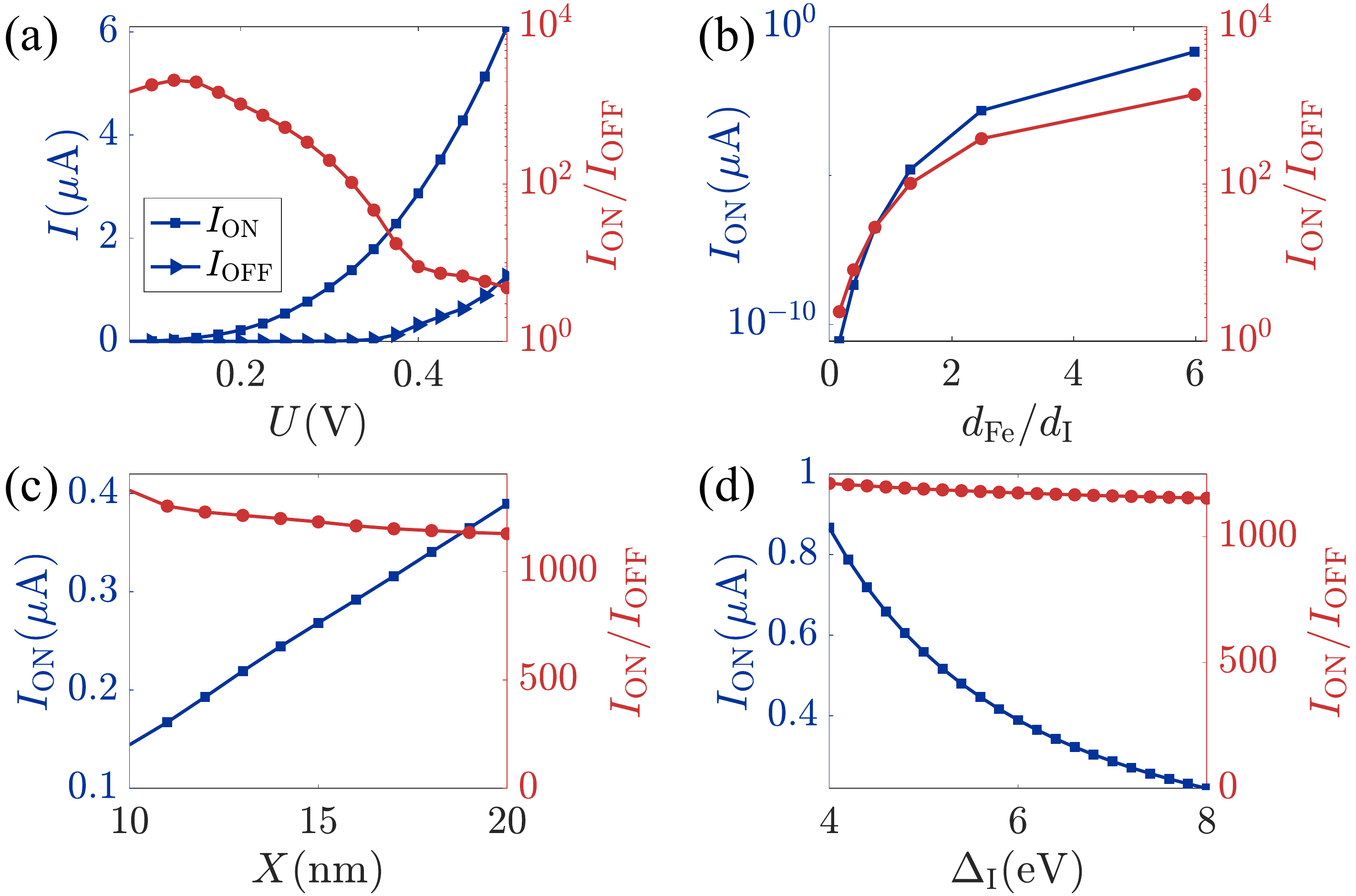}
\caption{Tunneling current computed from Landauer's formula Eq.~\eqref{eq:lan} as function of different parameters. $ X=Y=10\mathrm{nm} $, $ d_{\rm FE}=6\mathrm{nm} $, $ d_{\rm I}=1\mathrm{nm} $ and $ U=0.18\mathrm{V} $ if not specified. (a) Bias voltage $ U $; (b) Ferroelectric/insulator thickness ratio $d_{\rm FE}/d_{\rm I}$ (total thickness fixed to be $ Z=7\mathrm{nm} $); (c) Device width $ X $; (d) Insulator band gap $ \Delta_{\rm I} $.  }
\label{fig:current}
\end{figure}

Due to the complicated shape of the potential, the transmission probability in Eq.~\eqref{eq:lan} is computed numerically using Kwant package \cite{1367-2630-16-6-063065} based on the discretized version Eq.~\eqref{eq:schro}. Both tunneling current contributions from the electrons and the holes are taken into account. The results are summarized in Fig.~\ref{fig:current}. Since the magnitude of the tunneling current and the ON/OFF ratio are two quantities that determine the sensitivity and the accuracy of the peripheral current measuring device, we mostly focus on them. 

We first discuss the voltage-current characteristic (Fig.~\ref{fig:current}(a)). When the reading voltage is very small, both the current of the ON state and the OFF state come from the tunneling. With increasing the voltage, there is first a threshold in the ON state, after which $ V>V_{\text{ON}} $ and the Ohm's law $ I\propto U $ governs. The threshold voltage of the OFF state $V_{\text{OFF}}$ is larger than that of the ON state. In Fig.~\ref{fig:current}(a), $ V_{\text{ON}}\approx 0.1\mathrm{V} $ and $ V_{\text{OFF}}\approx 0.4\mathrm{V} $. When the voltage is in between the two threshold voltages, i.e. $ V_{\text{ON}}<V<V_{\text{OFF}} $, a very large ON/OFF ratio decreases exponentially with increasing† the voltage. In order to maximize the ON/OFF ratio, it is important for the bias voltage to be within this ``sweet spot''. 

From a device design point of view, the ON/OFF ratio and the size of the ``sweet spot'' $ V_{\text{OFF}}/V_{\text{ON}} $ can be enhanced by increasing the $d_{\rm FE}/d_{\rm I}$ ratio, as is shown in Fig.~\ref{fig:current}(b). This can be understood from our intuitive argument before---The difference of the tunneling region between the ON state and the OFF state is the ferroelectric film. This result suggests that the thickness of the insulator film $ d_{\rm I} $ should be small, but still large enough to prevent the electric discharge between the electrodes. Note that for fixed total thickness, increasing $d_{\rm FE}/d_{\rm I}$ also increases the magnitude of the ON state current significantly. The ON/OFF ratio can be as large as $\sim 1000$ for $ d_{\rm FE}/d_{\rm I}=6/1  $.

The magnitude of the ON state current after the threshold $ V>V_{\text{ON}} $ can be enhanced by simply increasing the width of the device in the $ x $ direction. As shown in Fig.~\ref{fig:current}(c), $ I_{\text{ON}} $ increases linearly with $ X $ because the number of modes per unit energy in the electrode also grows linearly. It is also possible to increase the current magnitude by decreasing the band gap of the insulator layer $ \Delta_{\rm I} $ (Fig.~\ref{fig:current}(d)). Note that for both approaches, the ON/OFF ratio is almost unaffected, implying an independent control of the ON state current and the ON/OFF ratio. 

We emphasize that although Fig.~\ref{fig:current} is computed using the parameter of SnTe, the qualitative conclusion is independent of the material. The device design and the transport model is completely general, which are applicable to any nanoplates or even nanodots with in-plane ferroelectricity, where the band bending can be both upward or downward, located at the material edge or the ferroelectric domain wall, and the charge carrier can be both electrons or holes. 

\section{Conclusion}
In conclusion, we have proposed a new type of ferroelectric memory based on the in-plane polarization. Compared with the DRAM or the out-of-plane polarization based FeRAM, our design has advantages including non-volatility, non-destructive reading operation, faster reading and writing operation, and the greater tunability of the tunneling current and the ON/OFF ratio. Our design is based on the ferroelectric thin film with in-plane polarization component that is switchable by an external electric field. Compared with FTJ, the in-plane polarization is much more robust in thin films. The 2D nature of the material, i.e. the weak van der Waals interlayer couplings, makes the device easy to fabricate. In particular, boundary effect caused by the lattice mismatching can be reduced greatly if the device is fabricated by stacking 2D materials \cite{Wang2015,Fei2018}.

A wide family of materials, for example, the IV-VI semiconductors XY, where X=Ge,Sn,Pb and Y=S,Se,Te, along with their alloys (for example, Pb$_{x}$Sn$_{1-x}$Te and Pb$_{x}$Sn$_{1-x}$Se) and superlattices (for example, PbTe/SnTe), are ideal candidates to realize our proposal \cite{2053-1583-4-1-015042,Wu2016,PhysRevLett.121.027601,Chang2016,PhysRevB.94.035304, PhysRevLett.117.246802}. Although SnTe nanoplates grown by molecular beam epitaxy (MBE) is not scalable at the moment, we believe that as more 2D ferroelectrics are being discovered, 2D ferroelectrics that can grow with uniform crystalline orientations can be found soon. For example, it is already found that In${}_2$Se${}_3$ has in-plane polarization and can be grown by both MBE and CVD (chemical vapor deposition) \cite{Ding2017,Zheng2018,doi:10.1021/acs.nanolett.7b04852,doi:10.1021/acs.nanolett.8b02688}. We hope our design can open up a new direction of ferroelectric non-volatile memories.

\begin{acknowledgments}
H.S. would like to thank Michal Papaj for helpful instructions on Kwant. This work is supported by the DOE Office of Basic Energy Sciences,
Division of Materials Sciences and Engineering under award de-sc0010526. J.L. acknowledges financial support from the Hong Kong Research Grants Council (Project No. ECS26302118). K.C. was funded by the Deutsche Forschungsgemeinschaft (DFG, German Research Foundation) – Project number PA 1812/2-1. L.F. is partly supported by the David and Lucile Packard Foundation. 
\end{acknowledgments}

\bibliography{FTRAM_Ref}

\widetext
\clearpage


\section*{Supplementary material}

# Supplemental Material for "In-Plane Ferroelectric Tunnel Junction"

Huitao Shen,[1] Junwei Liu,[2] Kai Chang,[3] and Liang Fu[1]

[1] *Department of Physics, Massachusetts Institute of Technology, Cambridge, Massachusetts 02139, USA*
[2] *Department of Physics, Hong Kong University of Science and Technology, Clear Water Bay, Hong Kong, China*
[3] *Max-Planck Institute of Microstructure Physics, Weinberg 2, 06120 Halle (Saale), Germany*

## I. QUANTUM TRANSPORT SIMULATION

As mentioned in the main text, we use two-terminal Landauer's formula to compute the tunneling current:

$$I = \frac{2e}{h} \int_{-\infty}^{\infty} T(E) \left[ f_L(E) - f_R(E) \right] dE, \tag{1}$$

where $T(E)$ is the transmission probability and $f_{L/R}(E) = \left[ e^{-(E - \mu_{L/R})/k_B T} + 1 \right]^{-1}$ is the Fermi-Dirac distribution function. $\mu_L$ and $\mu_R$ are the chemical potentials of the left and right electrode respectively. Here the factor of 2 is due to the spin degeneracy. In the following, we always set $\mu_L = 0$ as the reference. At zero temperature, $f_{L/R}(E)$ becomes the step function and Eq.(1) reduces to

$$I = \frac{2e}{h} \int_0^{\mu_R} T(E) dE = \frac{2e}{h} \int_0^U T(E) dE, \tag{2}$$

where $U \equiv \mu_R - \mu_L$ is the voltage bias.

Because of the complicated shape of the potential, we exploit the numerical method to compute the transmission probability, where we first discretize the Scrödinger equation. It is straightforward to show that the discretized version of

$$-\frac{\hbar^2}{2m^*} \left[ \frac{\partial^2 \psi}{\partial x^2} + \frac{\partial^2 \psi}{\partial y^2} + \frac{\partial^2 \psi}{\partial z^2} \right] + V(x, y, z)\psi = 0, \tag{3}$$

is

$$(6t + V(x, y, z))\psi(x, y, z) - t\left[ \psi(x + a, y, z) + \psi(x - a, y, z) + \right.$$
$$\left. \psi(x, y + a, z) + \psi(x, y - a, z) + \psi(x, y, z + a) + \psi(x, y, z - a) \right] = 0, \tag{4}$$

where $t = \hbar^2/(2m^* a^2)$. $a$ is the lattice constant in the discretized lattice and $a \to 0$ reduces Eq. (4) to (3). The discretization is a good approximation if $\lambda_F/a \gg 1$, where $\lambda_F$ is the Fermi wavelength.

The geometry of the system is the shown in Fig. 1. The size of the reading electrode is $X \times Y$. The thickness of the ferroelectric film and the insulator film are $d_{\text{FE}}$ and $d_{\text{I}}$ respectively. The total thickness (electrodes not included) is defined as $Z \equiv d_{\text{FE}} + d_{\text{I}}$.

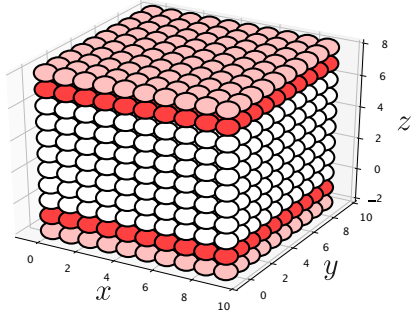

FIG. 1. Geometry of the system simulated in Kwant.

The potential barrier of the ferroelectric film can be modeled by an exponential potential and that of the insulator can be modeled simply by a square potential. Both electrons and holes can contribute to the tunneling current. The

TABLE I. Effective potentials for electrons and holes in different regions and states. Here we assume $\mu_{FE}, \mu_I < 0$ and $|\Delta_{FE}| > |\mu_{FE}|$, $|\Delta_I| > |\mu_I|$.

| | Ferroelectric Region | Insulator Region |
|---|---|---|
| $V_h$ | $\max(-\alpha e^{-y/\lambda} - \mu_{FE}, 0)$ | $-\mu_I$ |
| $V_e$ | $\max(\alpha e^{-y/\lambda} + \mu_{FE} + \Delta_{FE}, 0)$ | $\mu_I + \Delta_I$ |

TABLE II. Fitted band bending $V = \alpha e^{-y/\lambda} + c$ from STM measurements for SnTe thin films with 1 ML taken from Ref. [1]. Here the substrate is metallic graphene. Note that the fitting parameter $c$ is not the conduction band minimum (CBM) but the energy of the quantum well state.

| Upward | | | Downward | | |
|---|---|---|---|---|---|
| $\alpha(V)$ | $\lambda(nm)$ | $c(V)$ | $\alpha(V)$ | $\lambda(nm)$ | $c(V)$ |
| 0.198 | 4.85 | 1.57 | -0.329 | 6.12 | 1.57 |

effective potentials of these two tunneling channels are summarized in Table. I. $\mu_{FE}/\Delta_{FE}$ and $\mu_I/\Delta_I$ are the chemical potential/band gap of the ferroelectric and the insulator respectively.

In the following, we use the parameter of thin SnTe film for the ferroelectric region and that of monolayer h-BN for the insulator region. The band bending parameters are in Table. II. The band gap of the ferroelectric film and the insulator is chosen to be $\Delta_{FE} = 1.6$eV [1] and $\Delta_I = 6$eV respectively [2]. In principle, there exists boundary effect at the interfaces of the electrode and the ferroelectric. Since the parameters used in our calculation are taken from experimental measurements, this effect has already been taken into account.

The chemical potential of the insulator is set to be $\mu_I = -\Delta_I/2 = -3$eV respectively. The chemical potential of the ferroelectric is $\mu_{FE} = -0.1$eV for Fig. 3 in the main text. This also corresponds to Fig. 1(a) in the main text, where the main current contribution is from holes. The electrons are almost impossible to tunnel due to the high potential barrier in the ferroelectric region. An opposite scenario where the main charge carrier is the electron is discussed in the next section.

For a typical semiconductor, $m^*/m_e$ is usually between $0.01 - 0.1$ where $m_e$ is the free electron mass and $\lambda_F \sim 1 - 10$nm. In the following calculation, we use parameters for SnTe $m^*/m_e = 0.1$ [3]. For the geometry, we choose $X = Y = 10$nm. $d_{FE}$ and $d_I$ are variables that are to be tuned. Note that 1 ML SnTe is about 0.6nm thick [1]. In the simulation, we take $a = 1$nm [4].

## II. SIMULATION WITH ELECTRONS AS MAIN CHARGE CARRIERS

In the last section, the chemical potential is near the valence band of the ferroelectric and the main current contribution is from holes. In this section, we consider an opposite scenario where the main current contribution is from electrons.

To achieve this scenario, we take $\mu_{FE} = -1.4$eV, which corresponds to Fig. 1(b) in the main text. All other parameters are exactly the same as those in Fig. 3 in the main text. The results are shown in Fig. 2. Note that here the $I_{ON}$ is computed from the downward band bending instead of upward. Except for some quantitative difference, the results here are almost identical to Fig. 3 in the main text.

It is helpful to separate the current contribution from the hole channel and the electron channel in order to distinguish the two scenarios more clearly. The contribution of the tunneling current from the hole and the electron channel for both cases with upward band bending are plotted in Fig. 3. It is obvious that when the chemical potential is near the valence (conduction) band, the tunneling current mostly comes from holes (electrons). This is consistent with the schematic band bending in Fig. 1 in the main text.

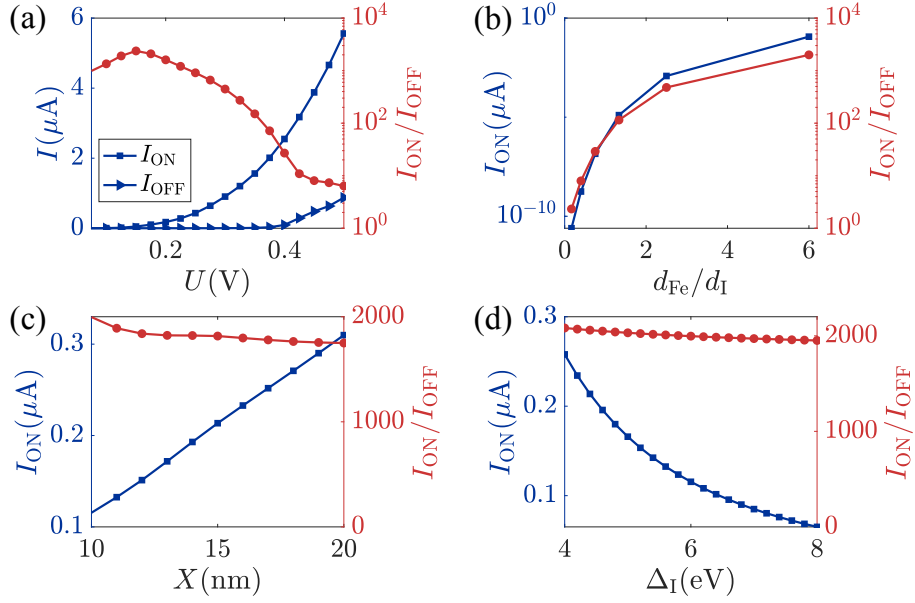

FIG. 2. Tunneling current computed from Landauer's formula Eq. (1) as function of different parameters. All parameters are the same as those in Fig. 3 in the main text except for the chemical potential $\mu_{\text{FE}}$.

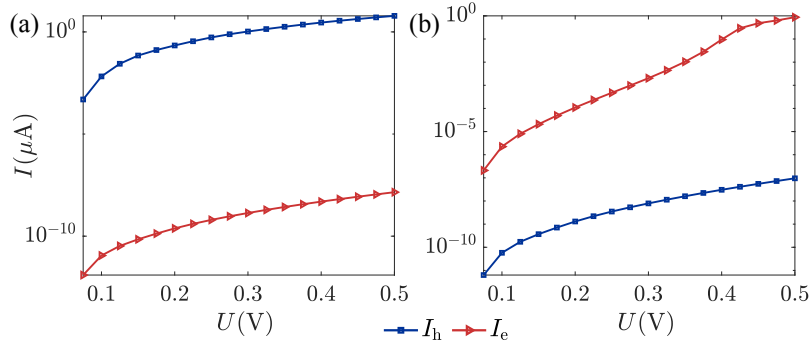

FIG. 3. Tunneling current from hole and electron channels when the chemical potential is $\mu_{\text{FE}} = -0.1$eV (a) and $-1.4$eV (b). The band is bending upward.

---


\end{document}